# Elastic, electronic, bonding, and optical properties of WTe$_2$ Weyl semimetal: A comparative investigation with MoTe$_2$ from first principles

B. Rahman Rano[1], Ishtiaque M. Syed[1], S. H. Naqib[2,*]

[1]Department of Physics, University of Dhaka, Dhaka-1000, Bangladesh

[2]Department of Physics, University of Rajshahi, Rajshahi-6205, Bangladesh

*Corresponding author e-mail: salehnaqib@yahoo.com

**ABSTRACT**

$T_d$-WTe$_2$ is a topological Weyl semimetal. WTe$_2$ in the orthorhombic structure is stable at room temperature. Elastic, electronic, bonding, and optoelectronic properties of WTe$_2$ have been investigated in detail in this work using the density functional theory. Elastic behaviour together with anisotropy indices of WTe$_2$ have been investigated for the first time. Bonding nature among the constituent atoms and electric field polarization dependent optical constants have also been explored for the first time. WTe$_2$ is elastically anisotropic; optical anisotropy on the other hand is low. The electronic band structure reveals quasi-linear dispersions along certain direction in the Brillouin zone with semi-metallic features. The Fermi level is located at a pseudogap separating bonding and anti-bonding density of states. The electronic effective mass tensor is predicted to be highly direction dependent. The energy dispersion is significantly weaker in the *c*-direction. The bonding in WTe$_2$ is an admixture of covalent and metallic bonds. Optoelectronic properties show strongly reflecting character over a wide band of photon energies. The compound is a strong absorber of ultraviolet radiation. The Debye temperature has been calculated from the elastic constants. We have compared all the calculated physical properties of WTe$_2$ with those of isostructural MoTe$_2$ Weyl semimetals. The properties of WTe$_2$ and MoTe$_2$ have been compared and contrasted. The calculated parameters of WTe$_2$ have also been compared with those already available in the literature. Very good agreements have been found.

**Keywords:** Density functional theory (DFT); Orthorhombic WTe$_2$; Weyl semimetal; Elastic constants; Band structure; Optical properties

I. Introduction

Hermann Weyl, in 1929, first suggested the existence of massless chiral fermions, known as the Weyl fermions [1]. These fermions are observed as low-energy quasiparticle excitations in Weyl semimetal (WSM) [2,3]. WSMs are identified by Fermi arc surface state at boundary by experimental techniques like angle resolved photoemission spectroscopy (ARPES). $WTe_2$ was predicted to be the first type-II WSM in 2015 [4]. They violate Lorentz invariance and have highly tilted Weyl cones on the Fermi surface. $WTe_2$ in its orthorhombic phase is a new generation of topological insulator holding many application related possibilities [5].

Electronic and crystallographic properties have long been studied in $WTe_2$ [6,7]. Moreover, since the theoretical prediction of type-II Weyl semi-metallic phase by Soluyanov *et al.* [4], there has been a flurry of research interest on this material. Fermi arcs were observed experimentally by different groups [8–12]. Pressure induced superconductivity and its origin in $WTe_2$ were also studied [13–15]. Then it has been reported that this pressure induced superconductivity is nearly isotropic [16]. Observation of extremely large, nonsaturating magnetoresistance by Ali *et al.* [17] is another reason of attracting interest in this compound. Quantum oscillation studies, NMR investigation and very recently mode-resolved reciprocal space mapping have been performed for $WTe_2$ WSM [18–20].

In our previous study, a comprehensive investigation of elastic and optical properties of orthorhombic $MoTe_2$ was done using DFT (density functional theory) [21]. We found that $MoTe_2$ has excellent reflecting characteristics with high level of machinability. $MoTe_2$ and $WTe_2$ are isostructural compounds, both belonging to the TMD (Transition-metal dichalcogenide) class. TMDs have the chemical formula $MX_2$, where $M$ is a transition metal (Mo, W, etc.) and $X$ is a chalcogenide atom (S, Se, and Te). $T_d$-$MoTe_2$ in its orthorhombic low temperature phase is a well-established type-II Weyl semimetal [22–24]. For $WTe_2$, the orthorhombic phase having Weyl state is stable at room temperature [25]. Moreover, the topological strength can be tuned in $WTe_2$ by doping with Mo [26]. Since we obtained some quality results for $MoTe_2$, we expected the same with $WTe_2$ for this comparative study. Importance of comparative study is hard to overstate. The similarities and the differences between $MoTe_2$ and $WTe_2$ arise due to the difference in the electronic configurations between Mo and W. Besides, how these two transition metals forms bonds with Te (a chalcogen) and among themselves determine the structural properties of $MoTe_2$ and $WTe_2$. Understanding of these aspects is instructive.

To the best of our knowledge, the elastic properties of bulk $WTe_2$ have not been investigated in any detail. Some optical properties of $WTe_2$ have been studied experimentally [27,28]. But their main focus was on temperature dependency and so the energy range was short. A thorough understanding of bulk properties i.e., elastic and optical properties are required to explore the possible applications of a material. In this work we have investigated the elastic and optical properties of semi-metallic $WTe_2$, complemented with the electronic band structure and energy density of states. In addition, we have compared these results with that of $MoTe_2$ from our earlier work [21]. Here we report the relative softness and higher optical reflectivity of WTe$_2$. Furthermore, anisotropic features are stronger in WTe$_2$. The comparative analysis of various other physical properties was carried out.

The rest of this work is organized as follows: In Section II, we shortly discussed the computational methodology and crystal structure. The results of our computations and their analyses are presented in Section III. Here we explored the structural and elastic properties, Debye temperature, electronic band structure, charge density distribution, bond population analysis and optical properties in different subsections. Finally, in Section IV, we summarized the key features of our investigations and drew some pertinent conclusions.

## II. Computational methodology and crystal structure

All the calculations presented in this work are performed using the DFT as implemented by the CAmbridge Serial Total Energy Package (CASTEP) [29]. The strength of DFT lies in the fact that in this formalism no attempt is made to compute the complex many-body wave functions. Instead the total energy of the system is expressed simply in terms of the electron density. This throws out the complexity of multi-dimensional electronic wave functions from the problem and brings density, a much simpler scalar to the forefront. Within DFT the ground state of a periodic solid is found by solving the Kohn-Sham equation [30]. Local density approximation (LDA) are used as the exchange-correlation functional [31] since it gives the best estimates to the lattice parameters for WTe$_2$. Vanderbilt-type ultra-soft pseudopotentials are utilized to take into account the interactions between the electrons and ions [32]. Density mixing electronic minimiser is used for the self-consistent calculations and Broyden Fletcher Goldfarb Shanno (BFGS) geometry optimization [33] algorithm is employed to optimize the crystal structure of WTe$_2$. To perform pseudo atomic calculations, the following valence electronic orbitals are used for W and Te atoms, respectively: W [5s$^2$ 5p$^6$ 5d$^4$ 6s$^2$], Te [5s$^2$ 5p$^4$]. k-point sampling within the reciprocal space (Brillouin zone) has been done with 7x5x3 regular mesh in the Monkhorst-pack grid scheme [34]. The cut-off energy for plane wave basis set has been set to 350 eV. This ensures satisfactory level of convergence of the energy during cell volume calculations. Geometry optimization has been performed using a self-consistent convergence limit of 10$^{-6}$ eV atom$^{-1}$ for the energy, 0.03 eV Å$^{-1}$ for the maximum force, 0.05 GPa for maximum stress, and 10$^{-3}$ Å for maximum atomic displacement.

At this point, we would like to mention that the surface electronic states carry the main topological features of WSMs and topological insulators. These novel surface electronic structures have their origin in the spin-orbit coupling (SOC). In this study we have not taken SOC into consideration. This is because our focus has been the bulk physical properties of WTe$_2$. In a number of prior investigations [35–40], we have demonstrated clearly that the bulk structural, elastic, bonding, optical, and thermo-physical properties of compounds belonging to different classes including topological semimetals and topological insulators, do not depend strongly on the SOC.

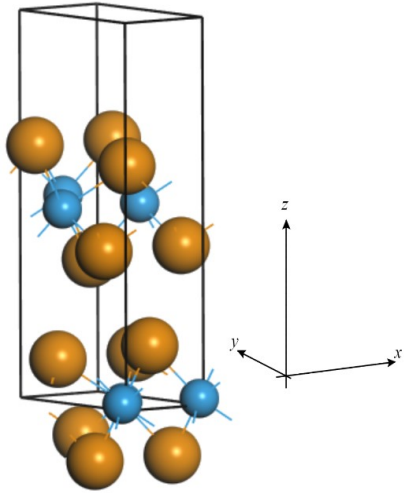

**FIG. 1.** Crystal structure of $T_d$-WTe$_2$. The blue spheres represent W atoms and the brown spheres represent Te atoms.

The crystal structure of WTe$_2$ is shown schematically in Fig. 1. Bulk WTe$_2$ crystallizes into the orthorhombic $T_d$ phase with $Pmn2_1$ (No. 31) space group. Atomic positions and lattice parameters of the crystal are fully optimized starting from the experimental data by Mar *et al*. [7]. The structure of WTe$_2$ is almost identical with $T_d$-MoTe$_2$. Both compounds have 4 W/Mo atoms and 8 Te atoms such that four formula units are incorporated in the unit cell. A slightly distorted octahedron is formed because of the displacement of W/Mo atoms from their ideal octahedral sites [5].

**TABLE I:** The structural parameters for $M$Te$_2$ ($M$ = Mo or W) obtained from the LDA calculations. The lattice parameters $a$, $b$, and $c$ are in Å and the unit cell volume ($V$) is in Å$^3$.

|  | $a$ | $B$ | $c$ | $V$ |
|---|---|---|---|---|
| WTe$_2$ | 3.464 | 6.209 | 13.750 | 295.74 |
| Experimental [7] | 3.477 | 6.249 | 14.018 | 304.60 |
| Theoretical [6] | 3.496 | 6.282 | 14.070 | 309.00 |
| MoTe$_2$ [21] | 3.458 | 6.297 | 13.294 | 289.48 |

The optimized structural parameters are listed in Table I together with some previously reported values. It can be seen that the calculated parameters agree well with the corresponding experimental values. Application of GGA (generalized gradient approximation) excessively overestimated the cell volume. The data by Mar *et al*. [7] was obtained at the lowest temperature. Our theoretically optimized geometry corresponds to the ground state (0 K). Also the use of LDA contracted the lattice due to localised nature of the trial wave functions. So our obtained values are slightly smaller than the experimental ones. The cell volume for MoTe$_2$ is somewhat lower than that of WTe$_2$. The ionic radii of Mo and W are quite similar; therefore, the difference in cell volume might arise due to the difference in the strength of atomic bonding in MoTe$_2$ and WTe$_2$. For completeness, all atoms occupy 2a Wyckoff positions with fractional coordinates of W: (0, 0.602952, 0.499863), (0, 0.042526, 0.015379); and Te: (0, 0.858725, 0.658542), (0, 0.645536, 0.111446), (0, 0.300531, 0.856578), and (0, 0.206037, 0.403482).

## III. Results and analysis

A. Elastic properties

Elastic properties of solids relate the mechanical and dynamical behavior under stress and create opportunities for industrial applications. Elastic constants are calculated using the 'stress-strain' method as contained in the CASTEP. Orthorhombic crystals have, by symmetry considerations, nine independent single crystal elastic constants in total and they are all given in Table II for WTe$_2$ and MoTe$_2$. Mechanical stability of a crystal system can be investigated using elastic constants [41]. For an orthorhombic structure the modified necessary and sufficient Born criteria are given by [42],

$$C_{11} > 0;\ C_{11}C_{22} > C_{12}^2$$

$$C_{11}C_{22}C_{33} + 2C_{12}C_{13}C_{23} - C_{11}C_{23}^2 - C_{22}C_{13}^2 - C_{33}C_{12}^2 > 0 \quad (1)$$

$$C_{44} > 0;\ C_{55} > 0;\ C_{66} > 0$$

Our calculated single crystal elastic constants satisfy these inequalities meaning that WTe$_2$ is stable mechanically.

**TABLE II:** The single crystal elastic constants ($C_{ij}$ in GPa) for $M$Te$_2$ ($M$ = Mo or W).

| ij | $C$ (WTe$_2$) | $C$ (MoTe$_2$ [21]) |
|---|---|---|
| 11 | 140.531 | 127.474 |
| 22 | 173.409 | 142.302 |
| 33 | 43.793 | 58.043 |
| 44 | 23.903 | 24.273 |
| 55 | 43.252 | 55.159 |
| 66 | 61.944 | 62.273 |
| 12 | 39.230 | 52.003 |
| 13 | 24.519 | 22.590 |
| 23 | 19.553 | 33.090 |

The three diagonal elastic constants $C_{11}$, $C_{22}$, and $C_{33}$ measure the capability of the crystal to resist tensile stress along the $a$, $b$ and $c$ axes, respectively. For WTe$_2$, $C_{33}$ is very small compared to $C_{11}$ and $C_{22}$. This indicates that the crystal is more compressible in the $c$-direction than the $a$- and $b$-directions defining the basal plane, reflecting the layered feature of the compound. Bonding strength in the *ab*-plane is therefore stronger than that extending in the out-of-plane. The constants $C_{44}$, $C_{55}$, and $C_{66}$ determine the resistance of the crystal against shear. Small $C_{44}$ indicates that WTe$_2$ is unable to resist shear deformation in (100) plane. This would constitute the prominent elastic failure mode for WTe$_2$. The off-diagonal shear components ($C_{12}$, $C_{13}$, and $C_{23}$) are due to the resistance to volume conserving orthorhombic distortions. $C_{23}$, which has the lowest value, describes a uniaxial strain along the crystallographic *c*-direction to a functional stress component along the crystallographic *b*-direction. For MoTe$_2$, $C_{13}$ has the lowest value, indicating a fundamental difference in the directional bonding characters between WTe$_2$ and MoTe$_2$.

Elastic moduli for polycrystalline aggregates can be calculated from single crystal elastic constants [43]. In Table III, the calculated polycrystalline bulk modulus (*B*), shear modulus (*G*), Pugh's ratio (*B/G*), Young's modulus (*E*), Poisson's ratio (*v*), and machinability index ($\mu_M$) for both WTe$_2$ and MoTe$_2$ are listed. The polycrystalline bulk and shear moduli are determined using the Voigt approximation and Reuss approximation [44,45]. These schemes give respectively the theoretical upper and lower bound of the elastic moduli. The Hill's approximation employs arithmetic mean of these two limits and closely represents the true polycrystalline constants [46].

**TABLE III:** The isotropic bulk modulus (*B* in GPa) and shear modulus (*G* in GPa) for polycrystalline *M*Te$_2$ (*M* = Mo or W) obtained from the single crystal elastic constants using Voigt, Reuss and Hill's approximations. The Pugh ratio (*B/G*), Young's modulus (*E* in GPa), Poisson's ratio (*v*), and the machinability index ($\mu_M$) are calculated from Hill's approximation.

| Compound | $B_R$ | $B_V$ | $B_H$ | $G_R$ | $G_V$ | $G_H$ | B/G | E | Y | $\mu_M$ |
|---|---|---|---|---|---|---|---|---|---|---|
| WTe$_2$ | 38.911 | 58.260 | 48.585 | 34.515 | 44.115 | 39.315 | 1.236 | 92.889 | 0.181 | 2.033 |
| MoTe$_2$ [21] | 48.309 | 60.354 | 54.332 | 35.971 | 43.017 | 39.494 | 1.380 | 95.373 | 0.207 | 2.238 |

Compared to many other metallic binary solids [39], the elastic moduli of WTe$_2$ are small, indicating its soft nature. Since *B* > *G*, the mechanical failure in WTe$_2$ should be dominated by shear component. The calculated value of *B*, *G* and *E* for WTe$_2$ is slightly lower than MoTe$_2$ indicating that WTe$_2$ is less hard. This is a consequence of weaker bonding strength in WTe$_2$. This weaker bonding strength is perhaps responsible for somewhat larger cell volume of WTe$_2$ in comparison to MoTe$_2$. Pugh's ratio and Poisson's ratio can separate the failure mode (ductility and brittleness) of solids with critical values of 1.75 and 0.26, respectively [47,48]. If the obtained value is less than the critical value then the material is predicted to be brittle. So both Pugh's ratio and Poisson's ratio for WTe$_2$ indicate that it is brittle in nature like its MoTe$_2$ counterpart. The value of Poisson's ratio is around 0.10 for covalent compounds and for metallic bonding, this value is around 0.33. Thus WTe$_2$ has a mixture of metallic and covalent bonding. Same conclusions can be drawn for MoTe$_2$. The bulk modulus to $C_{44}$ ratio is known as the machinability index [49]. A high value of $\mu_M$ for WTe$_2$ corresponds to easy machinability in the field of materials engineering.

**TABLE IV:** The bulk modulus ($B_{relax}$ in GPa) and its upper bound ($B_{unrelax}$ in GPa), bulk modulus along the orthorhombic crystallographic axes $a$, $b$, $c$ ($B_a$, $B_b$, $B_c$) and $\alpha$, $\beta$ for $M$Te$_2$ ($M$ = Mo or W).

| Compound | $B_{relax}$ | $B_{unrelax}$ | $B_a$ | $B_b$ | $B_c$ | $\alpha$ | $\beta$ |
|---|---|---|---|---|---|---|---|
| WTe$_2$ | 38.954 | 58.260 | 353.411 | 347.801 | 50.083 | 1.016 | 7.056 |
| MoTe$_2$ [21] | 47.974 | 60.354 | 223.966 | 483.623 | 69.871 | 0.463 | 3.205 |

The directional bulk moduli for a single crystal can also be calculated from the independent elastic constants [43]. We have tabulated these moduli in Table IV. $B_{relax}$ is the single crystal isotropic bulk modulus. Its value is the same as the one obtained from Reuss approximation. $B_{unrelax}$ gives the upper bound to the bulk modulus and is exactly the same as the one obtained from Vogit approximation for WTe$_2$. $B_c$ is very small compared to $B_a$ and $B_b$. This indicates that the bonding is weaker in $c$-direction which is consistent with the calculated elastic constants. $\alpha$ and $\beta$ are the relative change of the $b$ and $c$ axis as a function of the deformation of the $a$ axis. The values of $\alpha$ and $\beta$ of WTe$_2$ are almost double of those for MoTe$_2$. Such large difference in these parameters is interesting and requires further investigation.

**TABLE V:** The shear anisotropic factors $A_1$, $A_2$, $A_3$, and $A_G$ (in %), $A_B$ (in %) and compressibility anisotropy factors $A_{Ba}$, $A_{Bc}$ for $M$Te$_2$ ($M$ = Mo or W).

| Compound | $A_1$ | $A_2$ | $A_3$ | $A_G$ | $A_B$ | $A_{Ba}$ | $A_{Bc}$ |
|---|---|---|---|---|---|---|---|
| WTe$_2$ | 0.707 | 0.971 | 1.052 | 0.122 | 0.199 | 1.016 | 0.144 |
| MoTe$_2$ [21] | 0.692 | 1.645 | 1.503 | 0.089 | 0.111 | 0.463 | 0.145 |

Elastic anisotropy is very common in most of the crystals in nature and the study of it is important, especially for systems with layered structure. The estimated elastic anisotropy factors for both WTe$_2$ and MoTe$_2$ are listed in Table V. These factors are calculated using previously developed formalisms for orthorhombic systems [43,50,51]. Deviation from unity for $A_i$ (i = 1, 2, 3) measures the degree of elastic anisotropy. $A_1$ is the shear anisotropic factor for the {100} shear planes between the <011> and <010> directions. $A_2$ and $A_3$, which are very close to 1, are the factors for the {010} shear planes between the <101> and <001> directions and the {001} shear planes between the <110> and <100> directions, respectively. $A_B$ and $A_G$ are percentage anisotropies in compressibility and shear, respectively. These two indices are zero for an elastically isotropic crystal and 1 for the largest possible anisotropy. The levels of anisotropy are closely matching in both compressibility and shear for WTe$_2$ as well as MoTe$_2$. $A_{Ba}$ and $A_{Bc}$ are the compressibility anisotropies of the bulk modulus along the $a$ axis and $c$ axis with respect to the $b$ axis. Once again we have found $A_{Ba}$ of WTe$_2$ to be significantly higher than that for MoTe$_2$.

B. Debye temperature

Debye temperature ($\theta_D$) is an important lattice dynamical parameter correlated to many thermo-physical properties, such as specific heat, elastic constants, and melting temperature. $\theta_D$ sets the boson energy cut-off in Cooper pairing for the phonons involved in superconductors. The vibrational excitations arise entirely from acoustic modes at low temperatures. Hence, at low temperatures, specific heat measurements and calculation from the elastic constants give the same value of $\theta_D$. In this study, we have estimated $\theta_D$ from the averaged sound velocity and crystal density ($\rho$) using the following equation [52]:

$$\theta_D = \frac{h}{k}\left[\frac{3n}{4\pi}\left(\frac{N_A \rho}{M}\right)\right]^{1/3} v_m \qquad (2)$$

Here $h$ is the Planck's constant, $k$ is the Boltzmann constant, $N_A$ is the Avogadro's number, $\rho$ is the density, $M$ is the molecular weight, $n$ is the number of atoms in the molecule, and $v_m$ is the mean sound velocity. The longitudinal ($v_l$) and transverse ($v_t$) sound velocities of polycrystalline WTe$_2$ are obtained using the shear ($G$) and the bulk ($B$) modulus [43]. All the calculated values for WTe$_2$ along with MoTe$_2$ are listed in Table VI. As can be seen from Eqn. 2, $\theta_D$ decreases with increasing molecular weight. This is the main reason why WTe$_2$ has smaller Debye temperature compared to MoTe$_2$. Also, the lower the value of $\theta_D$, the softer is the material. So WTe$_2$ is softer than MoTe$_2$, which is consistent with the elastic moduli calculations.

**TABLE VI:** The density ($\rho$ in g/cm$^3$), longitudinal, transverse, average elastic wave velocity ($v_l$, $v_t$, $v_m$ in m/s), and the Debye temperature ($\theta_D$ in K) from the average elastic wave velocity obtained from polycrystalline elastic modulus.

| Compound | $\rho$ | $v_l$ | $v_t$ | $v_m$ | $\theta_D$ |
|---|---|---|---|---|---|
| WTe$_2$ | 9.863 | 3200.1 | 1996.5 | 2199.8 | 225.07 |
| MoTe$_2$ [21] | 8.054 | 3644.7 | 2214.4 | 2446.6 | 252.06 |

C. Electronic band structure and electronic energy density of states

The bulk electronic band structure along some high symmetry directions in the Brillouin zone (BZ) are depicted in Fig. 2 for both WTe$_2$ and MoTe$_2$. Fermi level, $E_F$ is set at 0 eV. Fig. 2 shows that the valence and conduction bands barely cross $E_F$ and the overlap between them is very small. This indicates the semi-metallic nature of the materials. Relatively larger band crossing can be observed in MoTe$_2$ which perhaps leads to better charge transport in this WSM. The bands running along $c$ axis in the k-space ($\Gamma$-Z, T-Y and X-U) are almost non-dispersive implying that the effective masses of charge carriers are high in these directions. A quasi-linearly dispersive conduction band weakly crosses $E_F$ along the Z-T direction for both the compounds. One major difference is noticed along the X-U direction in the momentum space. For MoTe$_2$, there is a definite crossing of the Fermi level, whereas for WTe$_2$, the valence and conduction bands are well separated with the conduction band lying ~ 0.30 eV above the Fermi level. The overall features of the band profile in this study are fairly analogous to those noticed in previous reports [6,26,27].

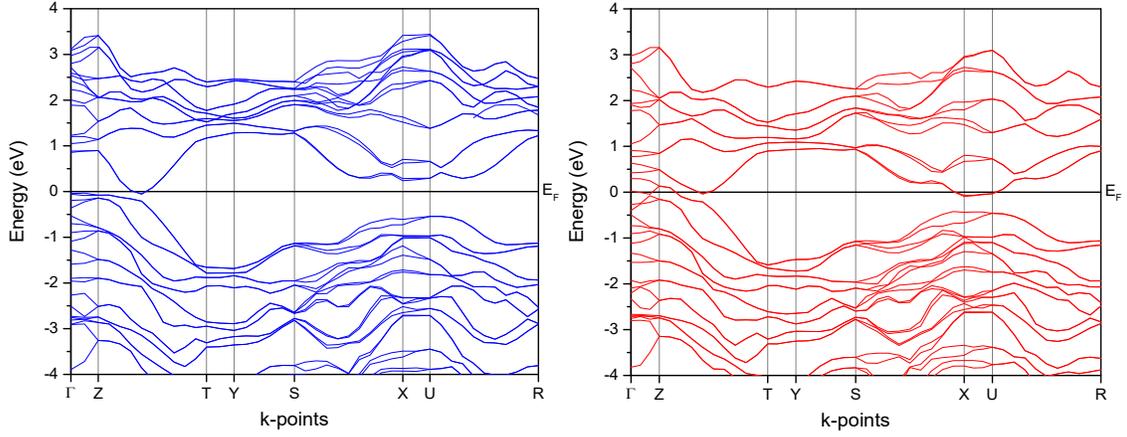

**FIG. 2.** The band structure of WTe$_2$ (left) and $T_d$-MoTe$_2$ (right) along the high symmetry directions in the BZ.

Fig. 3 shows the total density of states (TDOS) and the atomic orbital resolved partial density of states (PDOS) for WTe$_2$ and MoTe$_2$, as a function of energy, ($E$-$E_F$). Fermi level at 0 eV is represented by the vertical line. The pseudogap (deep valley) close the Fermi level indicates that the bonding is covalent with high electronic stability [43,53]. Pseudogap commonly exists between the bonding peak and anti-bonding peak, which are within 2 eV from Fermi level for WTe$_2$. So $E_F$ can be tuned to move across these peaks by chemical or mechanical means (e.g., doping or pressure). The formation of the covalent bond in WTe$_2$ could be due to the hybridization of the metal d and the chalcogen p orbitals since the main contributions in DOS come from the W-5d and Te-5p electronic orbitals. Above Fermi level, Te-5s and Te-5p orbitals form the conduction bands. The TDOS reaches a minimum, but does not go to zero near Fermi level. At $E_F$ it has a value of 3.2 states/eV-unit cell for WTe$_2$, which is slightly lower than MoTe$_2$ (4.55 states/eV-unit cell). So the carrier density of WTe$_2$ is expected to be lower than that of MoTe$_2$. We predict higher level of metallic conduction in MoTe$_2$ compared to WTe$_2$. The overall resemblance in the electronic properties for WTe$_2$ and MoTe$_2$ is owing to the similarity in the electronic configurations of the two metal atoms.

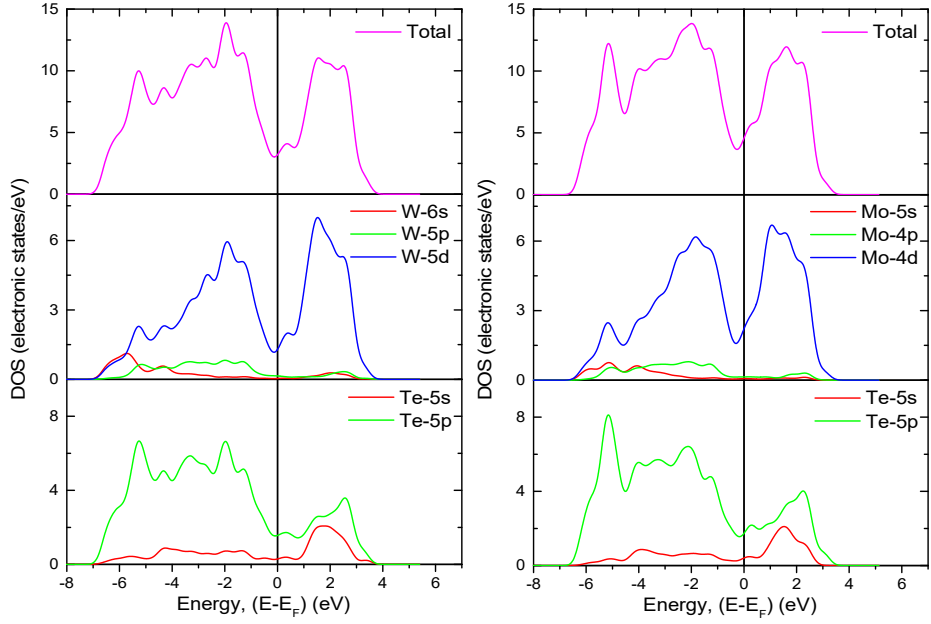

**FIG. 3.** Total and partial density of states for $WTe_2$ (left) and $T_d$-$MoTe_2$ (right).

D. Electronic charge density distribution

To visualize the bonding nature of the compounds, the charge density distribution in the (100) plane are studied. The charge density maps are illustrated in Fig. 4 for both $WTe_2$ and $MoTe_2$. Blue colour indicates high electron density and red colour means low electron density in this particular scale. Since the electronegativity of W/Mo and Te are comparable, no significant pull of electron density towards any of the atoms is noticed. Slight charge accumulation between W and Te refers to weak covalent bonding. Most of the bonding on this plane appears to be non-directional (metallic). So both of these compounds possess a mixture of metallic and covalent bonds which agree well with the Pugh and Poisson's ratio calculations presented in Section III A.

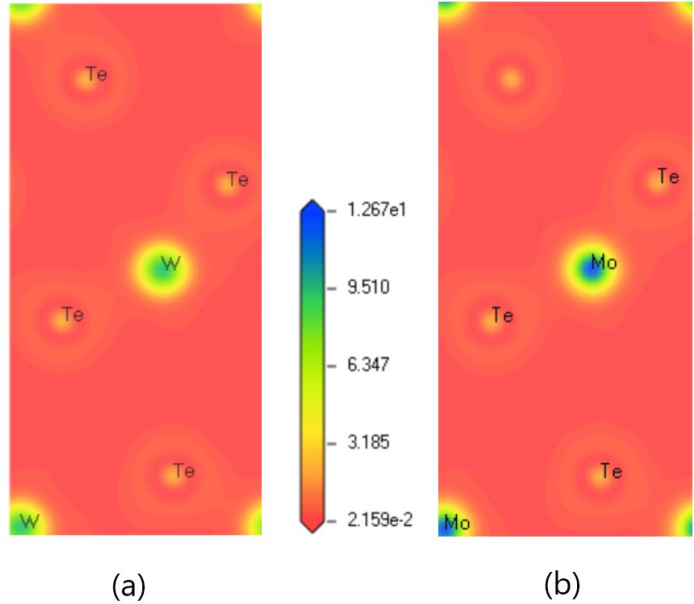

**FIG. 4.** Electronic charge density distribution map for $M$Te$_2$ (a) $M$ = W and (b) $M$ = Mo in the (100) plane.

E. Bond population analysis

To get a better understanding of electron density in various bonds, Mulliken population analysis (MPA) [54] and Hirshfeld population analysis (HPA) [55] have been performed. The calculated atomic populations are listed in Table VII. The charge spilling is low, which is a good indicator of reliable results. It is realized that in WTe$_2$, the charge transferred from Te to W is 0.80e according to Mulliken charge analysis; for MoTe$_2$ it is comparable, 0.90e. Our obtained d orbital charges agree well with the previous work [6]. The similarity in bonding character, according to MPA, of WTe$_2$ and MoTe$_2$ is apparent from Table VII. It is interesting to note that the Hirshfeld charge is very low for W, Te and Mo. Moreover, the signs of Hirshfeld charge are different for W (-0.06e) and Mo (0.04). This probably reflects the slight difference in the electronegativities of W (2.36) and Mo (2.16) in the Pauling scale. It is worth mentioning that MPA often overestimates the effective charge.

**TABLE VII:** Charge spilling parameter (%), orbital charges (electron), atomic Mulliken charge (electron), and Hirshfeld charge (electron) of $M$Te$_2$ ($M$ = Mo or W).

| Compound | Atoms | No. of ions | Charge spilling | s | P | d | Total | Mulliken charge | Hirshfeld charge |
|---|---|---|---|---|---|---|---|---|---|
| WTe$_2$ | W | 4 | 0.29 | 2.79 | 7.07 | 4.97 | 14.82 | -0.80 | -0.06 |
|  | Te | 8 |  | 1.74 | 3.86 | 0.00 | 5.60 | 0.40 | 0.03 |
| MoTe$_2$ [21] | Mo | 4 | 0.25 | 2.65 | 6.97 | 5.28 | 14.91 | -0.90 | 0.04 |
|  | Te | 8 |  | 1.69 | 3.85 | 0.00 | 5.55 | 0.45 | -0.02 |

Table VIII shows the bond population along with the bond lengths for WTe$_2$ and MoTe$_2$ in distorted octahedral coordination. No noteworthy interaction between the atoms should be observed since most of the overlap populations are close to zero. This is why these compounds are relatively soft. The positive values of the overlap population for the nearest neighbours indicate that they are bonded and negative values mean anti-bonded. Our calculated values of the bond lengths suggest that the strengths between *M-M* and *M*-Te are quite close and comparable. The metal-metal bond length is slightly greater for both compounds. The bond lengths we calculated show excellent agreement with the earlier work by Dawson *et al*. [6].

**TABLE VIII:** Calculated bond overlap populations and bond lengths (Å) for *M*Te$_2$ (*M* = Mo or W) WSMs.

| Bond | Bond number | Population (WTe$_2$) | Population (MoTe$_2$ [21]) | Length (WTe$_2$) | Length (WTe$_2$ [6]) | Length (MoTe$_2$ [21]) |
|---|---|---|---|---|---|---|
| *M-M* | 2 | 0.45 | 0.06 | 2.81 | 2.86 | 2.86 |
| *M*-Te | 2 | 0.34 | -0.06 | 2.69 | 2.71 | 2.68 |
| *M*-Te | 2 | 0.34 | -0.71 | 2.69 | 2.71 | 2.68 |
| *M*-Te | 2 | -0.43 | -0.07 | 2.70 | 2.71 | 2.69 |
| *M*-Te | 2 | -0.44 | -0.59 | 2.71 | 2.72 | 2.69 |
| *M*-Te | 2 | -0.03 | -0.14 | 2.78 | 2.81 | 2.76 |
| *M*-Te | 2 | 0.09 | -0.07 | 2.78 | 2.81 | 2.77 |
| *M*-Te | 2 | -0.49 | -0.67 | 2.80 | 2.81 | 2.79 |
| *M*-Te | 2 | -0.58 | -0.87 | 2.80 | 2.82 | 2.79 |

F. Optical properties

To investigate how the compound under investigation interacts with light, optical properties are studied. Optical parameters can be obtained by considering the photon induced electronic transitions. For the present study, a plasma frequency of 10 eV, 0.05 eV of damping in the Drude term and a Gaussian smearing of 0.5 eV are used. The imaginary part of the complex dielectric function, $\varepsilon_2(\omega)$, is obtained from the transitions between occupied and unoccupied electronic energy states, weighted by the corresponding matrix elements. This approach is provided by CASTEP supported formula, expressed as:

$$\varepsilon_2(\omega) = \frac{2e^2\pi}{\Omega\varepsilon_0}\sum_{k,v,c}|\langle\psi_k^c|\hat{u}.\vec{r}|\psi_k^v\rangle|^2\, \delta(E_k^c - E_k^v - E) \quad (3)$$

Here $\Omega$ is the volume of the unit cell, $\omega$ is the frequency of the incident photon, *e* is electronic charge, $\psi_k^c$ and $\psi_k^v$ are the quantum states of electrons in the conduction and valence bands, respectively, with a momentum given by $(h/2\pi)k$. Conservation of energy and momentum during the transition is ensured by the delta function. The real and imaginary parts of the dielectric constant describe a causal response and are linked by Kramers-Kronig transform. So this transformation is implemented in CASTEP to obtain the real part, $\varepsilon_1(\omega)$ of the dielectric function from the corresponding imaginary part, $\varepsilon_2(\omega)$. All the other optical parameters can be calculated from these two parts of the complex dielectric function [56].

Figs. 5 and Figs. 6 show the real and the imaginary parts of the dielectric constants, $\varepsilon_1(\omega)$ and $\varepsilon_2(\omega)$, real part of refractive index $n(\omega)$, extinction coefficient $k(\omega)$, real and imaginary parts of the optical conductivity $\sigma_1(\omega)$ and $\sigma_2(\omega)$, reflectivity $R(\omega)$, the absorption coefficient $\alpha(\omega)$, and the loss function $L(\omega)$ for $WTe_2$ and $MoTe_2$, respectively. All the energy dependent optical parameters are estimated for incident photon energies up to 20 eV. Electric field polarization vectors are taken along [100], [010] and [001] directions for both compounds.

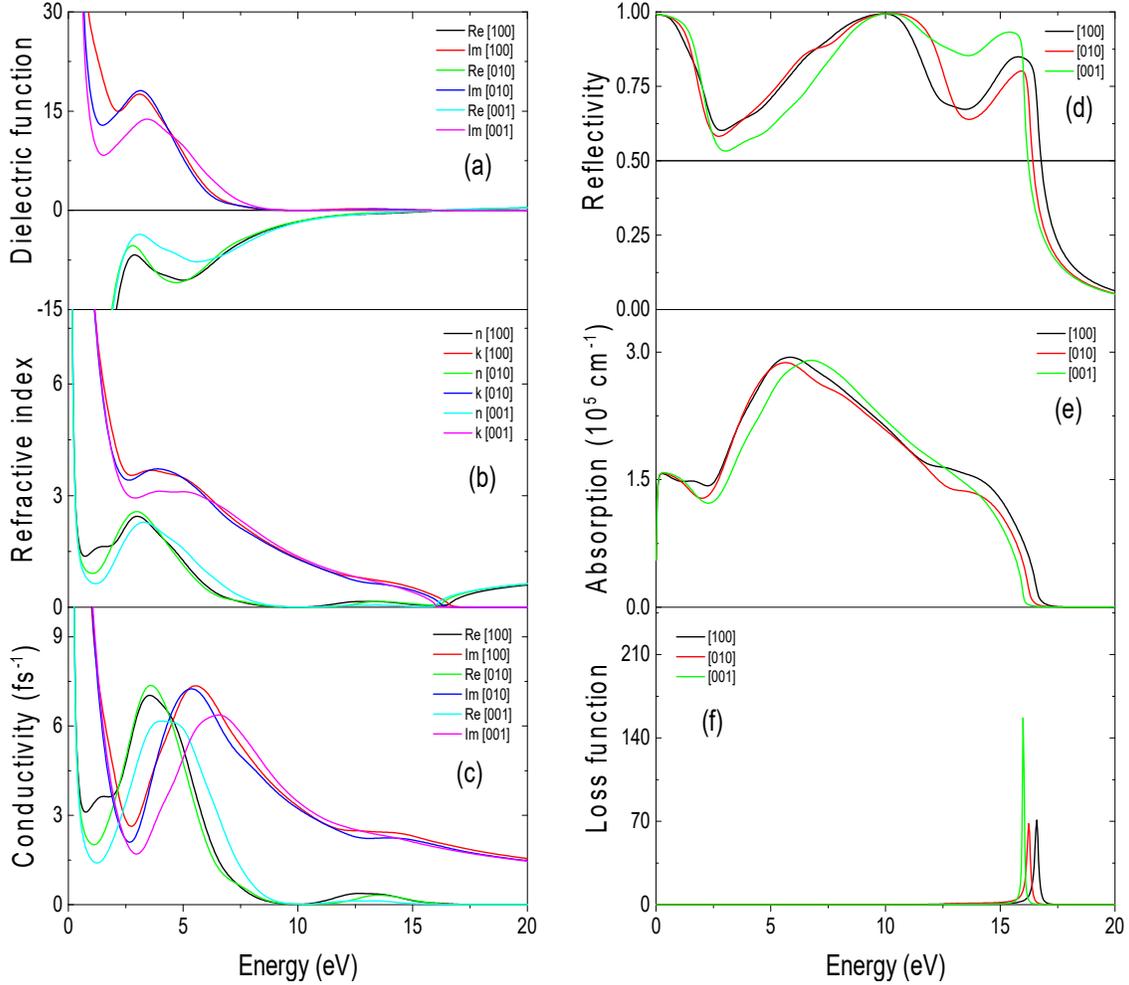

**FIG. 5.** The frequency dependent (a) dielectric function (real & imaginary parts), (b) refractive index (real & imaginary parts), (c) optical conductivity (real & imaginary parts), (d) reflectivity, (e) absorption coefficient, and (f) loss function of $WTe_2$ for different electric field polarization directions.

Optical response of the crystal can be described by the dielectric function at different photon energies. It can be seen from Fig. 5(a) that $\varepsilon_1(\omega)$, for all polarization directions, becomes zero from below at around 16 eV. $\varepsilon_2(\omega)$ also flattens to zero at and after this incident energy. For $WTe_2$, 16 eV is the plasma frequency above which the material becomes transparent and optical character becomes insulator-like. A broad peak at around 4 eV is seen for both real and imaginary parts. This could be owing to the electronic transitions between the bonding and the anti-bonding peak seen in the TDOS spectra [Fig. 3]. The real part of the refractive index is high in the visible region [see Fig. 5(b)]. The material under study is not transparent and therefore, the refractive index is not purely real. The imaginary part of the refractive index, also known as the extinction coefficient, is related with the absorption coefficient. At plasma edge, the real part becomes nonzero and the imaginary part falls to zero. Optical conductivities are nonzero at zero incident energy as can be seen from Fig. 5(c). This is an indication of the metallic nature of $WTe_2$ which is entirely in accord to the electronic band structure and DOS calculations.

The reflectivity spectra shown in Fig. 5(d) do not go below 50% and then fall sharply at 16 eV, the plasma edge. This means that $WTe_2$ can be used in the fabrication of optoelectronic devices where wide band high reflectivity is required. Reflectivity is almost 100% in the infrared region which was also found by experimental studies [27,28]. $R(\omega)$ is minimum near 3 eV where $n(\omega)$ is maximum. A finite value of $\alpha(\omega)$ in Fig. 5(e) at 0 eV is again indicating that $WTe_2$ is metallic in nature. The absorption peak is in the ultraviolet region meaning that the material can be a good UV ray absorber. After the plasma frequency, $\alpha(\omega)$ is zero as expected. The loss function, depicted in Fig. 5(f), describes the energy lost by an electron passing through the material. The loss peak, found at 16 eV, is due to the plasma resonance [57].

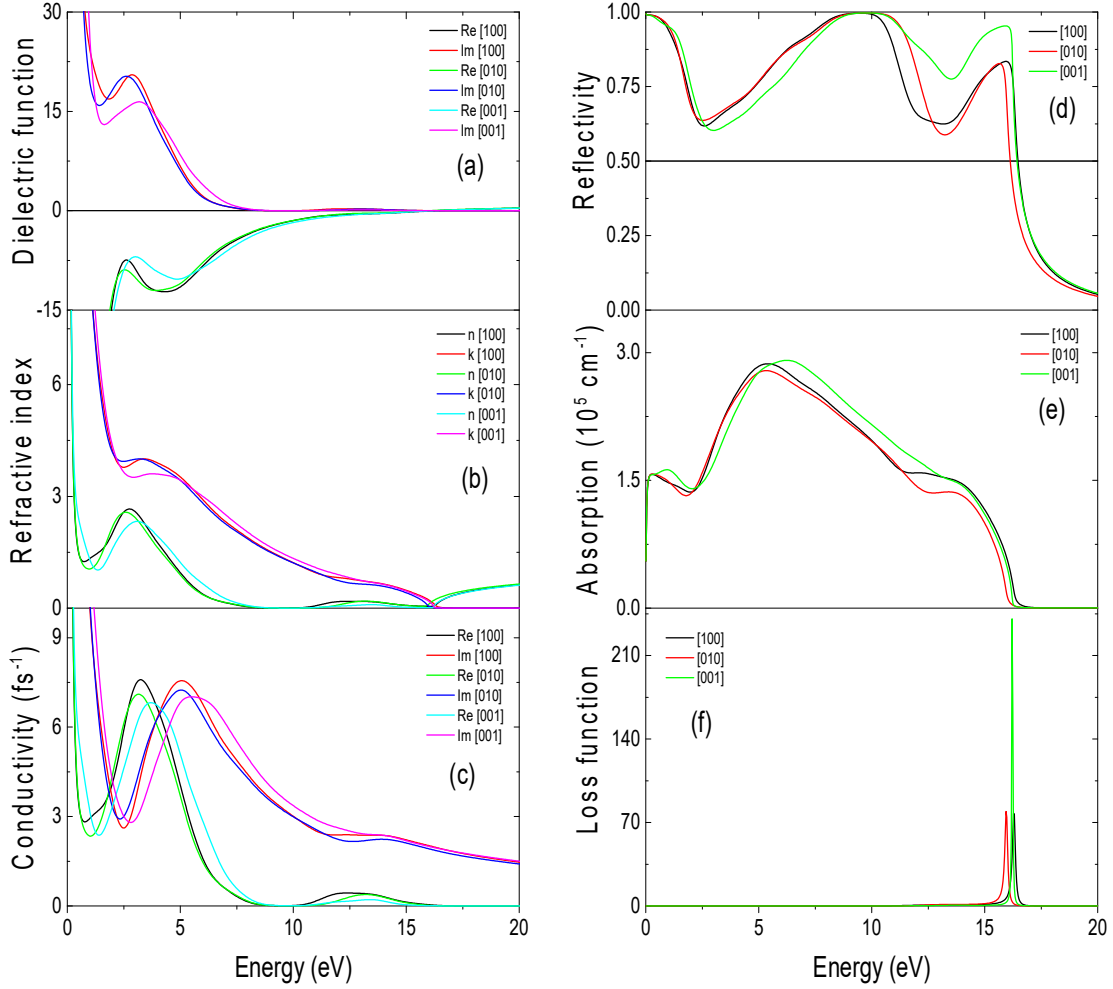

**FIG. 6.** The frequency dependent (a) dielectric function (real & imaginary parts), (b) refractive index (real & imaginary parts), (c) optical conductivity (real & imaginary parts), (d) reflectivity, (e) absorption coefficient, and (f) loss function of $MoTe_2$ for different electric field polarization directions.

Figs. 6 depict different optical parameters for $MoTe_2$. It can be seen that apart from optical anisotropy, other features are almost identical to $WTe_2$. Anisotropic feature is stronger in $WTe_2$ when the electric field is along *c*-direction. For both materials the refractive index and the reflectivity is high in the visible region. The plasma frequency is around 16 eV as well. Though for $WTe_2$ the height of loss peak is lower compared to $MoTe_2$. However, both $WTe_2$ and $MoTe_2$ [21] are good candidates for optoelectronic device applications.

## IV. Conclusions

Elastic, electronic, bonding and optical properties of WTe$_2$ WSM were investigated via first-principles technique. All the computed properties of WTe$_2$ were compared and contrasted with those of MoTe$_2$ WSM in the isostructural orthorhombic $T_d$ phase. Similar to MoTe$_2$, WTe$_2$ is an elastically anisotropic compound brittle in nature. The polycrystalline elastic moduli of WTe$_2$ WSM are slightly lower compared to those for MoTe$_2$. The calculated unit cell volume of WTe$_2$, on the other hand, is larger than the cell volume of MoTe$_2$. Both these features imply that bonding strength is relatively low in WTe$_2$. Both WTe$_2$ and MoTe$_2$ WSMs possess high and comparable values of machinability index which augurs well for possible applications in the industrial sector. The Debye temperature of WTe$_2$ is significantly lower compared to that of MoTe$_2$. Thus the lattice thermal conductivity of WTe$_2$ is expected to be lower. It also implies that the crystal lattice of WTe$_2$ is softer in comparison to MoTe$_2$.

Electronic band structures of both WTe$_2$ and MoTe$_2$ WSMs exhibit high level of resemblance with semi-metallic characters and agree very well with previous experimental studies [6,26,27]. Band structures of both the compounds show quasi-linear energy dispersion along the Z-T direction of the BZ. This implies very high mobility of electrons in this particular band. The electronic density of states at the Fermi level is higher in MoTe$_2$. The degree of metallic conduction is expected to be higher in MoTe$_2$ in comparison to the WTe$_2$ WSM. Both the WSMs possess significant pseudogap at the $E_F$. Such a feature implies that the band structure of both the materials can be engineered relatively easily to modify the DOS value at the Fermi level by pressure or alloying with suitable dopants.

The bonding characters have been explored using the charge density mapping, Mulliken and Hirshfeld bond population analyses. Clear indications of an admixture of covalent and metallic bonding have been found. The directional covalent bonding results in the brittleness of WTe$_2$ and MoTe$_2$ WSMs.

The refractive indices of both the WSMs are found to be very high in the visible region meaning that WTe$_2$ and MoTe$_2$ might be a good candidate for optical displays. The reflectivity spectra suggest that both WTe$_2$ and MoTe$_2$ compounds will be good reflecting materials over a wide band of energies encompassing from infrared to ultraviolet regions. The gross features of the recently measured experimental optical conductivity and reflectivity [27,28] are in good agreement with the theoretical results presented in this study. Both WTe$_2$ and MoTe$_2$ WSMs are found to be very good absorber of ultraviolet radiation. The peak structures in the optical parameters are consistent with the underlying DOS profiles. Dielectric constants, optical absorption, and photoconductivity spectra reaffirm the metallic character (absence of band gap) of WTe$_2$ and MoTe$_2$, as found from the electronic band structure calculations. The loss peak appears at around the same energy (~ 16 eV) for both the compounds of interest. Therefore, the plasma frequencies for WTe$_2$ and MoTe$_2$ are expected to be almost identical.

We hope that the results presented in this work will inspire both experimentalists and theorists to study these interesting WSMs in further details in near future.

**Acknowledgements**

S. H. N. acknowledges the research grant (1151/5/52/RU/Science-07/19-20) from the Faculty of Science, University of Rajshahi, Bangladesh, which partly supported this work.

**Data availability**

The data sets generated and/or analyzed in this study are available from the corresponding author on reasonable request.

**References**

[1]   H. Weyl, Proc. Natl. Acad. Sci. **15**, 323 (1929).

[2]   S. Y. Xu, I. Belopolski, N. Alidoust, M. Neupane, G. Bian, C. Zhang, R. Sankar, G. Chang, Z. Yuan, C. C. Lee, S. M. Huang, H. Zheng, J. Ma, D. S. Sanchez, B. K. Wang, A. Bansil, F. Chou, P. P. Shibayev, H. Lin, S. Jia, and M. Z. Hasan, Science (80-. ). **349**, 613 (2015).

[3]   B. Q. Lv, H. M. Weng, B. B. Fu, X. P. Wang, H. Miao, J. Ma, P. Richard, X. C. Huang, L. X. Zhao, G. F. Chen, Z. Fang, X. Dai, T. Qian, and H. Ding, Phys. Rev. X **5**, 31013 (2015).

[4]   A. A. Soluyanov, D. Gresch, Z. Wang, Q. Wu, M. Troyer, X. Dai, and B. A. Bernevig, Nature **527**, 495 (2015).

[5]   W. Tian, W. Yu, X. Liu, Y. Wang, and J. Shi, Materials (Basel). **11**, (2018).

[6]   W. G. Dawson and D. W. Bullett, J. Phys. C Solid State Phys. **20**, 6159 (1987).

[7]   A. Mar, S. Jobic, and J. A. Ibers, J. Am. Chem. Soc. **114**, 8963 (1992).

[8]   F. Y. Bruno, A. Tamai, Q. S. Wu, I. Cucchi, C. Barreteau, A. De La Torre, S. McKeown Walker, S. Riccò, Z. Wang, T. K. Kim, M. Hoesch, M. Shi, N. C. Plumb, E. Giannini, A. A. Soluyanov, and F. Baumberger, Phys. Rev. B **94**, (2016).

[9]   Y. Wu, D. Mou, N. H. Jo, K. Sun, L. Huang, S. L. Bud'Ko, P. C. Canfield, and A. Kaminski, Phys. Rev. B **94**, (2016).

[10]   C. Wang, Y. Zhang, J. Huang, S. Nie, G. Liu, A. Liang, Y. Zhang, B. Shen, J. Liu, C. Hu, Y. Ding, D. Liu, Y. Hu, S. He, L. Zhao, L. Yu, J. Hu, J. Wei, Z. Mao, Y. Shi, X. Jia, F. Zhang, S. Zhang, F. Yang, Z. Wang, Q. Peng, H. Weng, X. Dai, Z. Fang, Z. Xu, C. Chen, and X. J. Zhou, Phys. Rev. B **94**, (2016).

[11] J. Sánchez-Barriga, M. G. Vergniory, D. Evtushinsky, I. Aguilera, A. Varykhalov, S. Blügel, and O. Rader, Phys. Rev. B **94**, (2016).

[12] B. Feng, Y. H. Chan, Y. Feng, R. Y. Liu, M. Y. Chou, K. Kuroda, K. Yaji, A. Harasawa, P. Moras, A. Barinov, W. Malaeb, C. Bareille, T. Kondo, S. Shin, F. Komori, T. C. Chiang, Y. Shi, and I. Matsuda, Phys. Rev. B **94**, (2016).

[13] D. Kang, Y. Zhou, W. Yi, C. Yang, J. Guo, Y. Shi, S. Zhang, Z. Wang, C. Zhang, S. Jiang, A. Li, K. Yang, Q. Wu, G. Zhang, L. Sun, and Z. Zhao, Nat. Commun. **6**, (2015).

[14] X. C. Pan, X. Chen, H. Liu, Y. Feng, Z. Wei, Y. Zhou, Z. Chi, L. Pi, F. Yen, F. Song, X. Wan, Z. Yang, B. Wang, G. Wang, and Y. Zhang, Nat. Commun. **6**, (2015).

[15] P. Lu, J. S. Kim, J. Yang, H. Gao, J. Wu, D. Shao, B. Li, D. Zhou, J. Sun, D. Akinwande, D. Xing, and J. F. Lin, Phys. Rev. B **94**, (2016).

[16] Y. T. Chan, P. L. Alireza, K. Y. Yip, Q. Niu, K. T. Lai, and S. K. Goh, Phys. Rev. B **96**, (2017).

[17] M. N. Ali, J. Xiong, S. Flynn, J. Tao, Q. D. Gibson, L. M. Schoop, T. Liang, N. Haldolaarachchige, M. Hirschberger, N. P. Ong, and R. J. Cava, Nature **514**, 205 (2014).

[18] Z. Zhu, X. Lin, J. Liu, B. Fauqué, Q. Tao, C. Yang, Y. Shi, and K. Behnia, Phys. Rev. Lett. **114**, (2015).

[19] A. O. Antonenko, E. V. Charnaya, M. K. Lee, L. J. Chang, J. Haase, S. V. Naumov, A. N. Domozhirova, and V. V. Marchenkov, Phys. Solid State **61**, 1979 (2019).

[20] P. Hein, S. Jauernik, H. Erk, L. Yang, Y. Qi, Y. Sun, C. Felser, and M. Bauer, Nat. Commun. **11**, (2020).

[21] B. R. Rano, I. M. Syed, and S. H. Naqib, J. Alloys Compd. **829**, 154522 (2020).

[22] Y. Sun, S. C. Wu, M. N. Ali, C. Felser, and B. Yan, Phys. Rev. B - Condens. Matter Mater. Phys. **92**, 161107 (2015).

[23] L. Huang, T. M. McCormick, M. Ochi, Z. Zhao, M. T. Suzuki, R. Arita, Y. Wu, D. Mou, H. Cao, J. Yan, N. Trivedi, and A. Kaminski, Nat. Mater. **15**, 1155 (2016).

[24] K. Deng, G. Wan, P. Deng, K. Zhang, S. Ding, E. Wang, M. Yan, H. Huang, H. Zhang, Z. Xu, J. Denlinger, A. Fedorov, H. Yang, W. Duan, H. Yao, Y. Wu, S. Fan, H. Zhang, X. Chen, and S. Zhou, Nat. Phys. **12**, 1105 (2016).

[25] H. J. Kim, S. H. Kang, I. Hamada, and Y. W. Son, Phys. Rev. B **95**, (2017).

[26] T. R. Chang, S. Y. Xu, G. Chang, C. C. Lee, S. M. Huang, B. K. Wang, G. Bian, H. Zheng, D. S. Sanchez, I. Belopolski, N. Alidoust, M. Neupane, A. Bansil, H. T. Jeng, H. Lin, and M. Zahid Hasan, Nat. Commun. **7**, (2016).

[27] S. I. Kimura, Y. Nakajima, Z. Mita, R. Jha, R. Higashinaka, T. D. Matsuda, and Y. Aoki, Phys. Rev. B **99**, 195203 (2019).

[28] A. J. Frenzel, C. C. Homes, Q. D. Gibson, Y. M. Shao, K. W. Post, A. Charnukha, R. J. Cava, and D. N. Basov, Phys. Rev. B **95**, (2017).

[29] S. J. Clark, M. D. Segall, C. J. Pickard, P. J. Hasnip, M. I. J. Probert, K. Refson, and M. C. Payne, Zeitschrift Fur Krist. **220**, 567 (2005).

[30] W. Kohn and L. J. Sham, Phys. Rev. **140**, A1133 (1965).

[31] J. P. Perdew and A. Zunger, Phys. Rev. B **23**, 5048 (1981).

[32] D. Vanderbilt, Phys. Rev. B **41**, 7892 (1990).

[33] T. H. Fischer and J. Almlöf, J. Phys. Chem. **96**, 9768 (1992).

[34] H. J. Monkhorst and J. D. Pack, Phys. Rev. B **13**, 5188 (1976).

[35] M. I. Naher and S. H. Naqib, J. Alloys Compd. **829**, (2020).

[36] M. A. Afzal and S. H. Naqib, arXiv:2005.13393.

[37] M. I. Naher and S. H. Naqib, arXiv:2005.10590.

[38] F. Parvin and S. H. Naqib, J. Alloys Compd. **780**, 452 (2019).

[39] M. I. Naher, F. Parvin, A. K. M. A. Islam, and S. H. Naqib, Eur. Phys. J. B **91**, (2018).

[40] F. Parvin and S. H. Naqib, Chinese Phys. B **26**, 106201 (2017).

[41] M. Born, Math. Proc. Cambridge Philos. Soc. **36**, 160 (2013).

[42] F. Mouhat and F. X. Coudert, Phys. Rev. B - Condens. Matter Mater. Phys. **90**, 224104 (2014).

[43] P. Ravindran, L. Fast, P. A. Korzhavyi, B. Johansson, J. Wills, and O. Eriksson, J. Appl. Phys. **84**, 4891 (1998).

[44] W. Voigt, *Lehrbuch Der Kristallphysik* (Teubner Leipzig, 1928).

[45] A. Reuss, ZAMM - J. Appl. Math. Mech. / Zeitschrift Für Angew. Math. Und Mech. **9**, 49 (1929).

[46] R. Hill, Proc. Phys. Soc. Sect. A **65**, 349 (1952).

[47] S. F. Pugh, London, Edinburgh, Dublin Philos. Mag. J. Sci. **45**, 823 (1954).

[48]  G. N. Greaves, A. L. Greer, R. S. Lakes, and T. Rouxel, Nat. Mater. **10**, 823 (2011).

[49]  Z. Sun, D. Music, R. Ahuja, and J. M. Schneider, Phys. Rev. B - Condens. Matter Mater. Phys. **71**, 193402 (2005).

[50]  C. M. Kube, AIP Adv. **6**, 95209 (2016).

[51]  S. I. Ranganathan and M. Ostoja-Starzewski, Phys. Rev. Lett. **101**, 55504 (2008).

[52]  O. L. Anderson, J. Phys. Chem. Solids **24**, 909 (1963).

[53]  J. H. Xu, T. Oguchi, and A. J. Freeman, Phys. Rev. B **35**, 6940 (1987).

[54]  R. S. Mulliken, J. Chem. Phys. **23**, 1833 (1955).

[55]  F. L. Hirshfeld, Theor. Chem. Acc. **44**, 129 (1977).

[56]  S. Azam, S. A. Khan, R. Khenata, S. H. Naqib, A. Abdiche, Uğur, A. Bouhemadou, and X. Wang, Mol. Phys. **118**, (2020).

[57]  X. C. Ma, Y. Dai, L. Yu, and B. B. Huang, Light Sci. Appl. **5**, (2016).

**Competing Interests**

The authors declare no competing interests.

**CRediT authorship contribution statement**

**B. Rahman Rano:** Methodology, Software, Investigation, Formal analysis, Draft writing

**Ishtiaque M. Syed:** Supervision, Reviewing the draft manuscript

**S. H. Naqib:** Conceptualization, Supervision, Writing the manuscript, Formal analysis